\documentclass[prl,aps,twocolumn,showpacs,amsmath]{revtex4-1}
\usepackage[breaklinks, colorlinks=true]{hyperref}
\usepackage{graphicx}
\usepackage{bm}

\newcommand{\unit}[1]{\,\mathrm{#1}}
\newcommand{\vkick}{v_{\text{kick}}}
\newcommand{\nl}{\nonumber\\}
\newcommand{\muCSE}{\mu_{\text{CSE}}}
\newcommand{\be}{{\bm e}}
\newcommand{\bj}{{\bm j}}
\newcommand{\bk}{{\bm k}}
\newcommand{\br}{{\bm r}}
\newcommand{\bB}{{\bm B}}
\newcommand{\me}{m_{\text{e}}}
\newcommand{\mue}{\mu_{\text{e}}}
\newcommand{\nB}{n_{\text{B}}}
\newcommand{\nsube}{n_{\text{e}}}
\newcommand{\Ye}{Y_{\text{e}}}
\newcommand{\calL}{\mathcal{L}}
\newcommand{\calM}{\mathcal{M}}

\begin{document}

\title{Pulsar Kick by the Chiral Anisotropy Conversion}
\date{\today}
\author{Kenji Fukushima}
\email{fuku@nt.phys.s.u-tokyo.ac.jp}
\author{Chengpeng Yu}
\email{yu.chengpeng@nt.phys.s.u-tokyo.ac.jp}
\affiliation{Department of Physics, The University of Tokyo, 7-3-1 Hongo, Bunkyo-ku, Tokyo 113-0033, Japan}

\begin{abstract}
  We discuss a novel mechanism for the proto-neutron star acceleration assisted by the chiral separation effect which induces an axial vector current in a dense medium.  We consider the process of neutrinos scattering off the background axial vector current of electrons.  We show that anisotropy of either magnetic field or density in momentum space is essential for nonzero recoil and we call this mechanism the chiral anisotropy conversion.  Assuming a strong magnetic field $B \simeq 10^{12} \,\mathrm{T}$ and anisotropy by $\sim 10\%$, we find that the chiral anisotropy conversion can yield the velocity of order of typical pulsar kicks, i.e., $\vkick \gtrsim 1000\,\mathrm{km/s}$.
\end{abstract}
\maketitle

%%%%%%%%%%
\paragraph*{Introduction:}

Pulsars are neutron stars (NSs) emitting electromagnetic pulses and many are known to move at high speed with the mean velocity around $\vkick\simeq 400\unit{km/s}$ (see Ref.~\cite{Hobbs:2005yx} for a statistical study of 233 pulsar proper motions).  Interestingly, the top tail $\sim 15\%$ of the velocity distribution reaches as fast as $\gtrsim 1000\unit{km/s}$ which is orders of magnitude higher than their progenitor, i.e., supernovae~\cite{Long2022,Igoshev2021}.  This phenomenon referred to as the pulsar kick is not only challenging in theory, but also it may provide us with valuable information about properties of neutrinos~\cite{Ayala:2019sbt} and dense matter~\cite{Schmitt2005,Jiang2020}.  Moreover, the information about the NS star velocity distribution is relevant to the galaxy evolution~\cite{Chu2022,Kusenko2008}.

There are a number of models to explain the origin of the pulsar kicks.  The most conventional scenario is that high speed pulsars originate from the recoil of the anisotropic supernova explosion when they were born~\cite{Muller2019}; see Ref.~\cite{Lai:2003hm} for a review and Refs.~\cite{Muller2019,Nakamura2019,Powell2020} for some pulsar kick arguments.  The required anisotropy could be generated before the core collapse~\cite{Burrows:1995bb} and amplified later~\cite{Lai:1999hi}.  According to 3D hydrodynamic simulation, however, this model still has some limitations such as too small predicted velocity $\vkick\sim 100\unit{km/s}$~\cite{Muller2019,Nakamura2019,Powell2020,Metlitski2005,Schmitt2005}, too strong mass-velocity correlation not seen in observation~\cite{Muller2019,Nakamura2019} (which also causes a difficulty of explaining the pulsar kick of low-mass NSs with nearly symmetric ejecta~\cite{Gessner:2018ekd}), and the observed bimodal structure in the velocity distribution that cannot be reproduced in the model~\cite{Schmitt2005,Igoshev2021}.

\begin{figure}
    \includegraphics[width=0.7\columnwidth]{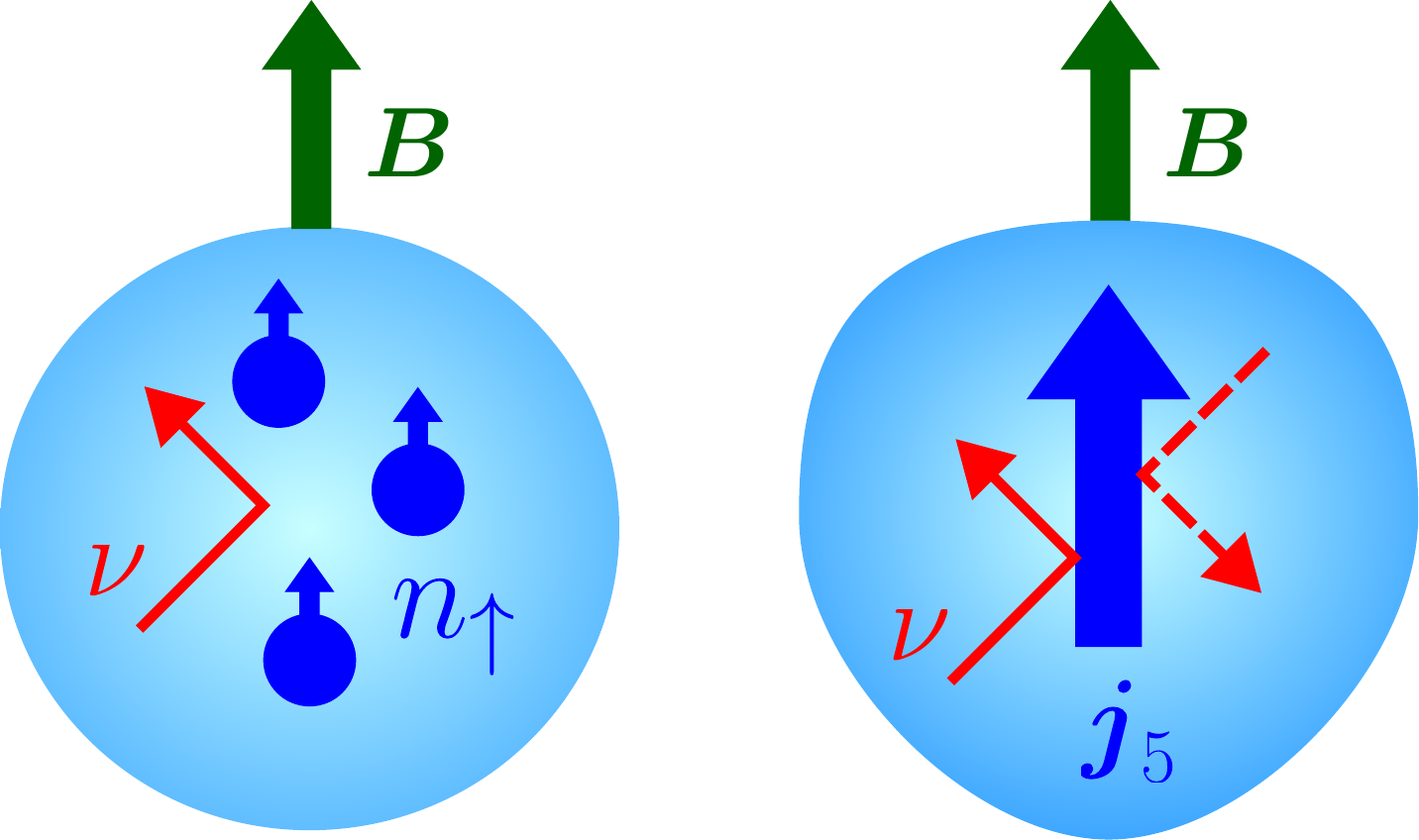}
    \caption{Schematic figures for the pulsar kick mechanisms.  (Left) Scattering between neutrinos and polarized nucleons under the magnetic field.  The parity violation in the weak interaction induces anisotropy.  (Right) Scattering between neutrinos and the background axial current }
    \label{fig:scheme}
\end{figure}

These limitations hint that the pulsar kicks should be driven by not only one simple mechanism but it may well be a consequence of composite effects.  One essential ingredient for additional kick is the strong magnetic field $\bB$.  Charged particles are polarized under $\bB$.  Because the weak interaction breaks parity symmetry, the neutrino scattering with polarized particles has a preferred direction and the anisotropic emission of neutrinos from the proto-neutron star (PNS) kicks the pulsars~\cite{Kusenko1996,Ayala2018}.  For this, the neutrino-nucleon scattering is the dominant process~\cite{Arras:1998mv} than the neutrino-electron scattering~\cite{1987SvAL...13..282C,1995ApJ...451..700V,Horowitz:1997mk} (see the left in Fig.~\ref{fig:scheme} for the schematic illustration).
%However, for a pulsar, most neutrinos are emitted as the result of cooling of the star in the first minute after its birth (the proto-neutron star (PNS) stage); at this stage, the temperature of the star is still very high ($T\sim 10\;\mathrm{MeV}$) \cite{Camelio2017}, so the neutrinos would scatter many times as they move across the crust of the star, and these scatterings would lead to the loss of ansiotropy \cite{Prakash2000,Adhya2017}; as a result, the magnetic field required by this type of model is relatively high \cite{Long2022,Ayala2018,Lai1998}.
Related to this, $\bB$ may induce gauged vortices in quark matter cause beaming of neutrino emission~\cite{Berdermann:2006rk}.

Recently, moreover, the anomaly induced pulsar kick has been discussed intensively.  The magnetic field gives rise to various exotic transport phenomena, such as the chiral magnetic effect (commonly called the CME) $\bj = \mu_5/(2\pi^2) e\bB$, the chiral separation effect (CSE) $\bj_5 = \mu/(2\pi^2) e\bB$, the chiral vortical effect (CVE) $\bj_5 = [(\mu^2+\mu_5^2)/(2\pi^2)+T^2/6] e\bm{\omega}$, etc.\ (see Refs.~\cite{Kharzeev:2015znc,Kamada:2022nyt} for reviews in nuclear physics and in astrophysics, respectively).  In Ref.~\cite{Metlitski2005} it is indicated that the mean free path of the induced current could be sufficiently long to penetrate from the core to the PNS surface.  Hence, the multiple scattering generally suppresses anisotropy~\cite{Prakash2000,Adhya2017}, but such suppression can be bypassed by the induced current that transfers the momentum to surface neutrinos.  Later on, related works along these lines~\cite{Charbonneau:2009ax, Kaminski2016, Shaverin2018} have attempted to quantify this mechanism.  More recently, an intriguing possibility was pointed out in Refs.~\cite{Yamamoto2023-1}.  The neutrino transport gives a recoil to electrons, and the resulting charge current is reminiscent of the CME\@.  Then, this effective CME induces the chiral plasma instability~\cite{Akamatsu2013} with simultaneous growth of $\bB$ and $\bj$, which enhances the pulsar kicks.  The predicted kick velocity is $\vkick\simeq  B/(10^9\mbox{-}10^{10}\unit{T}) \unit{km/s}$ for a certain initial condition.

We are proposing a novel mechanism as a robust result of the combination of anisotropy and chiral transport.  As mentioned above, the anisotropy is naturally expected before and during the core collapse, and we show that neutrinos scatter off the anisotropic background axial current $\bj_5$, which converts momentum anisotropy to the pulsar kick effectively (see the right in Fig.~\ref{fig:scheme}).

We emphasize qualitative differences of our mechanism from the preceding works.  In comparison to the scattering scenario under $\bB$, on the one hand, one might think that $\bj_5$ appears from the polarization, and thus the neutrino scattering with $\bj_5$ physically corresponds to the scattering with polarized nucleons/electrons.  This is partially correct, but as we discuss later with explicit calculations, constant $\bj_5$ is sensitive to parity-even processes and this is why external anisotropy should be coupled in our mechanism.  Regarding the chiral transport, on the other hand, the previous works~\cite{Kaminski2016,Yamamoto2021} assume a hydrodynamic regime in which neutrinos and electrons are equilibrated after scatterings and $\bj_5$ involves neutrinos as well.  In our case we consider the microscopic processes that can occur even before the hydrodynamic regime.

For the quantitative analysis we need the time evolution of the density distribution of electrons.  To this end we can refer to the state-of-the-art modeling of proto-neutron stars~\cite{Pons1999, Ofengeim2017, Raduta2021}, among which we specifically adopt the numerical results from Ref.~\cite{Pons1999}.  Because those simulations do not take account of the $\bj_5$ effect, we neglect the back-reaction in the present work.  For other back-reaction effects such as anisotropy by the self-energy, see Ref.~\cite{Yamamoto:2023okm}.  Apart from this approximation, we avoid theoretical uncertainty by utilizing the energy luminosity of emitted neutrinos instead of the internal neutrino distribution.
\vspace{1em}

%--- figure ---%

%%%%%%%%%%
\paragraph*{Formulation:}

% axial current
The axial current is the essential ingredient for our pulsar-kick mechanism.  To determine the CSE-induced contribution to the axial current, we should estimate the values of parameters in the following formula:
\begin{equation}
    \bm{j}_{5,e} = \frac{e}{2\pi^{2}}\muCSE(\mue, \me, T) \bB \,.
    \label{eq:j5_e0}
\end{equation}
We consider the electron contribution only in the above formula, while the baryons can make finite contributions to the axial current.  Due to the baryon mass much larger than the typical PNS temperature, $T\sim 10\unit{MeV}$, the baryon contributions are negligible.
We introduced the effective chemical potential including the mass effect:
\begin{equation}
    \muCSE(\mu,m,T) = \int_0^{\infty} \!\!dq\, \bigl[ f(q;\mu,m,T) \!-\! f(q;-\mu,m,T) \bigr]\,,
    \label{eq:CSE}
\end{equation}
where the integrand is the Fermi-Dirac distribution function, i.e., $f(q;\mu,m,T) = 1/[e^{(\sqrt{q^2+m^2}-\mu)/T}+1]$.  It is easy to confirm that $\muCSE(\mu,m=0,T)=\mu$.

The electron mass, $\me=0.511\unit{MeV}$, is a physical parameter, and we need to fix $\mue$, $T$, and $\bB$ as functions of the time $t$.  For this purpose we must rely on a PNS cooling model and in this work we adopt the results reported in Ref.~\cite{Pons1999}.  For related works based on the results from Ref.~\cite{Pons1999}, see Refs.~\cite{Villain2004,Camelio2017,Nakazato2019} for example.
From Ref.~\cite{Pons1999}  we took the PNS data for $M=1.6M_\odot$ in the models ``GM3np'', ``GM1np'', ``GM3npH'', and ``GM1npH'', respectively.  These are mean-field models in which protons and neutrons interact via $\omega$ and $\rho$ exchange.  They are different in the choice of coupling parameters (GM1 and GM3) and whether hyperons are included (np and npH).  We make use of the baryon number density, $\nB$, the temperature, $T$, the net electron concentration, $\Ye=\nsube/\nB$, and the total neutrino energy luminosity, $L_E$, as functions of $t$ where the supernova explosion sets $t=0$.  Specifically, we read the data from Figs.\ 9, 15, 16, 18 in Ref.~\cite{Pons1999}.

%We shall explain how to fix the electron chemical potential $\mue$.
%Under the assumption that we can treat electrons as a free gas, the following relation should hold:
%\begin{equation}
%    \nsube = \Ye \nB = 2\int\frac{d^3 q}{(2\pi)^3} \frac{1}{e^{(\sqrt{q^2+m_e^2}-\mu_e)/T}+1}\,.
%    \label{eq:n_mu}
%\end{equation}
Since $\Ye$ and $\nB$ are both given in Ref.~\cite{Pons1999}, we can numerically solve $\mue$ from $\nsube=\Ye \nB$.  For $T=0$, the solution is analytically found as $\mue|_{T=0} = \sqrt{(3\pi^2 \Ye\nB)^{2/3}+\me^2}$.
Even at $T\neq 0$, $\mue|_{T=0}$ is a reasonable approximation.
%At $T\neq 0$ we have numerically solved Eq.~\eqref{eq:n_mu} and verified that $\mue|_{T=0}$ reasonably approximates the results in the typical range of the PNS temperatures.
As for the magnetic field, $\bB$, our formulation requires some anisotropy in the combination of $\mue$ and $\bB$.  Therefore, for convenience, we keep $\bB=B_0\be_z$ constant for the moment and implement anisotropy into the axial current later.

% anisotropy
%According to the model simulation it is an acceptable assumption that the spatial dependence of $\mue$ and $T$ is characterized by one parameter, $M_B(r)$, which is the enclosed mass inside the PNS from the center up to the sphere radius $r$.
%In this way, we could deal with the radially inhomogeneous distribution, but we need to implement anisotropy breaking rotational symmetry.

%Since $M_\mathrm{en}$ is nothing but a function of $r$, the axial current distribution obtained from Eq.~(\ref{eq:j5_e0}) is isotropic, i.e., the axial current distribution is the same on all longitudes and latitudes of the PNS. However, under the real situation, the PNS could acquire anisotropic density and magnetic field profiles from the anisotropic supernova explosions. This anisotropy could be crucial for modeling the pulsar kick accurately. Therefore, we need to figure out a natural way to implement an anisotropic axial current distribution.

Suppose the dependence on the radial distance $r$ is fixed, we obtain $\bj_{5,e}(r)$, and its Fourier transform, $\tilde{\bj}_{5,e}(k)$.  Using the polar coordinates in momentum space, i.e., $\bk=k(\sin\theta\cos\varphi,\sin\theta\sin\varphi,\cos\theta)^T$, we can employ the spherical harmonics, $Y_l^m(\theta,\varphi)$, as the general basis.
%The deformed axial current distribution can be expressed as
%\begin{equation}
%    \tilde{\bj}_{5,e}(\bk) = \tilde{\bj}_{5,e}(k) \sum_{l=0}^{\infty} \sum_{m=-l}^{l} a_{l,m} Y_l^m(\theta,\varphi)\,,
%\end{equation}
Among $Y_l^m(\theta,\varphi)$, only the $m=0$ part is important due to axial symmetry.
%Because neutron stars rotate rapidly, the $\varphi$ dependence of the induced current should be averaged out, so that we can simply choose $m=0$.  Thus, $\tilde{\bj}_{5,e}(\bk)$ becomes
%\begin{equation}
%    \tilde{\bj}_{5,e}(\bk) = \tilde{\bj}_{5,e}(k) \sum_{l=0}^{\infty} \alpha_{l} P_l(\cos\theta)\,.
%\end{equation}
%Here, we denote $\alpha_l = \sqrt{(2l+1)/(4\pi)}\, a_{l,0}$ including the normalization.
Without loss of generality we can expand the spherical harmonics up to the first order, i.e.,
\begin{equation}
    \tilde{\bj}_{5,e}(\bk)
    = \tilde{\bj}_{5,e}(k) \bigl( 1 + \alpha_1\cos\theta+\cdots
    %\frac{1}{2}\alpha_2(3\cos^2\theta-1)+\cdots
    \bigr)\,.
    \label{eq:iso_and_aniso}
\end{equation}
This suffices for our present purpose of the order of magnitude demonstration.
%Since we are interested only in the order of magnitude, we keep the expansion up to $\alpha_1$ and drop higher-order angular dependence.

% cross-section

%--- figure ---%
%\begin{figure}
%    \includegraphics[width=0.8\columnwidth]{fig_ZW.pdf}
%    \caption{Two contributions to the scattering mediated by $Z$ (left) and $W$ (right).}
%    \label{fig:ZW}
%\end{figure}
%--- figure ---%

Hereafter, let us consider the process in the weak interaction involving neutrinos and $\tilde{\bj}_{5,e}$.  The interacting term in the effective Lagrangian density for the weak interaction density has two contributions from the neutral and the charged currents.
The neutral current represents $\nu_e e \to \nu_e e$ via the $Z$ exchange, i.e., $\calL_{\text{NC}} = -\bigl[\bar{e}\gamma_\mu (c_L^{\nu_e e} P_L + c_R^{\nu_e e} P_R) e\bigr]\, \bar{\nu}_e \gamma^\mu P_L \nu_e$
with $P_L=\frac{1}{2}(1-\gamma_5)$ and $P_R=\frac{1}{2}(1+\gamma_5)$.
Here, $c_L^{\nu_e e}-c_R^{\nu_e e}\simeq \sqrt{2}G_F$.
This form has direct coupling to the electron axial current and the the interaction induced by the background current is given by
$\calL_{\text{eff,NC}} = \frac{G_F}{2\sqrt{2}} (j_{5,e})_\mu \bar{\nu}_e \gamma^\mu (1-\gamma_5) \nu_e$.
Also, the effective Lagrangian involving the charged current via the $W^\pm$ exchange is identified from the Fierz transformation as
$\calL_{\text{CC}} = \frac{G_F}{\sqrt{2}} \bar{\nu}_e \gamma_\mu P_L e\,\bar{e}\gamma^\mu P_L\nu_e$.
We can extract the coupling to the electron axial current by taking the Fierz transformation back, leading to $\calL_{\text{eff,CC}}=2\calL_{\text{eff,NC}}$.  In total, we find the mean-field effective interaction as
\begin{equation}
    \calL_{\text{eff}} = \frac{3G_F}{2\sqrt{2}} (j_{5,e})_\mu \bar{\nu}_e\gamma^\mu(1-\gamma_5)\nu_e\,.
    \label{eq:Lag}
\end{equation}
The Fermi coupling is related to the Higgs condensate as $(\sqrt{2}G_F)^{-1/2}=246\;\mathrm{GeV}$.
The axial currents from protons and neutrons involving $u$ and $d$ quarks are negligible because the nucleon mass is larger than the PNS temperature by one order of magnitude.

We note that this effective interaction describes the scattering of neutrinos with the background axial current field, i.e., $\nu_e + j_{5,e} \to \nu_e$ and that of anti-neutrinos, i.e., $\bar{\nu}_e + j_{5,e} \to \bar{\nu}_e$.  We have neglected the pair production and annhilation processes such as $j_{5,e} \leftrightarrow \nu_e + \bar{\nu}_e$.  We can justify this treatment from the typical energy and time scales.
That is, the PNS cools down within the time of order of seconds, and the energy scale associated with time variation is extremely tiny, that is, $E\sim 1\,\text{s}^{-1}\sim 10^{-21}\text{MeV}$.  Therefore, we can regard the axial current as a static background.  The typical momentum scale should be characterized by the system size, and we can expect $p \sim 1/(10\unit{km})\sim 10^{-17}\unit{MeV}$.  These scales, $E \ll p$, are too small as compared to the QCD scale, and the energy-momentum conservation in effect prohibits the pair production and annihilation processes.

In this way we can just focus on the scattering processes.  We can estimate the amplitude for $\nu_e(k_1) + j_{5,e} \to \nu_e(k_2)$, where $k_1$ and $k_2$ are the four-momenta carried by the incoming and the outgoing neutrinos as
\begin{equation}
    i\calM(k_1,k_2) = i\frac{3G_F}{\sqrt{2}} \chi_L^\dagger(k_2)\bar{\sigma}^\mu\chi_L(k_1)\,(\tilde{j}_{5,e}(\bk_2-\bk_1))_\mu\,,
    \label{eq:amp}
\end{equation}
where $\chi_L$ is a solution of the Weyl equation for the neutrino satisfying $k_\mu \bar{\sigma}^\mu \chi_L = 0$.  The solution takes a form of $\chi_L(k)=(\sqrt{E_{\bk}-k^z}, e^{i\varphi(\bk)}\sqrt{E_{\bk}+k^z})^T$ with $E_{\bk}=|\bk|$ and $e^{i\varphi(\bk)}=(k^x+ik^y)/\sqrt{(k^x)^2+(k^y)^2}$.
Using the polar coordinates in momentum space, we can simplify the expression as $\chi_L(k)=\sqrt{2E_{\bk}}(\sin\frac{\theta_k}{2}, e^{i\varphi_k}\cos\frac{\theta_k}{2})$.  We choose the $z$-axis along $\bB\parallel \bj_{5,e}$.  Then, the coupled neutrino current is $\chi_L^\dag(k_2)\sigma^z\chi_L(k_1)=2\sqrt{E_1 E_2}[\sin\frac{\theta_2}{2}\sin\frac{\theta_1}{2}-e^{i(\varphi_1-\varphi_2)}\cos\frac{\theta_2}{2}\cos\frac{\theta_1}{2}]$.  
Here, for notation brevity, we introduced the notation, $E_1$, $\theta_1$, and $\varphi_1$ for the energy and the angular variables instead of $E_{\bk_1}$, $\theta_{\bk_1}$, and $\varphi_{\bk_1}$, and $E_2$, $\theta_2$, and $\varphi_2$ similarly.
The squared quantity is
$|\chi_L^\dag(k_2)\sigma^z\chi_L(k_1)|^2 = 2E_1 E_2[1+\cos\theta_1\cos\theta_2-\sin\theta_1\sin\theta_2\cos(\varphi_1-\varphi_2)]$.  In the same way we get the amplitude for $\bar{\nu}_e(k_1) + j_{5,e} \to \bar{\nu}_e(k_2)$ by
$\chi_L^\dag(k_2)\to \chi_L^\dag(k_1)$ and
$\chi_L(k_1)\to \chi_L(k_2)$ in Eq.~\eqref{eq:amp}.

%\begin{equation}
%    i\bar{\calM}(k_1,k_2) = i\frac{3G_F}{\sqrt{2}} \chi_L^\dagger(k_1)\bar{\sigma}^\mu\chi^L(k_2) (\tilde{j}_{5,e}(\bk_2-\bk_1))_\mu\,.
%\end{equation}
%Obviously, we find $|\bar{\calM}|^2=|\calM|^2$.
After all, the cross section associated with the scattering between (anti-)neutrinos and the electron axial current turns out to be
\begin{align}
    & d\sigma = d\bar{\sigma} = \frac{1}{2E_1} |\calM(k_{1},k_{2})|^{2} 2\pi\delta(E_2-E_1) \frac{d^3\bk_2}{2E_2(2\pi)^3} \nl
    & = \frac{9G_{F}^{2}}{16\pi^2}[1+\cos\theta_{1}\cos\theta_{2}-\sin\theta_{1}\sin\theta_{2}\cos(\varphi_{1}-\varphi_{2})] \nl
    &\quad \times |\tilde{j}_{5,e}^z(\bk_2-\bk_1,t)|^2 \, \delta(E_2-E_1) \, d^3\bk_2\,.
    \label{eq:scat-cross}
\end{align}

% acceleration
%Knowing the scattering cross-section, we can integrate over the phase space of incident and scattered neutrinos to compute the total momentum transfer from the axial current.  In this way we can estimate the pulsar kick acceleration based on momentum conservation.
%However, because of Eq.~(\ref{eq:scale_rel}), the $\tilde{j}_{5,e}^{z}(\bm{k}_{2}-\bm{k}_{1},t)$ in Eq.~(\ref{eq:scat-cross}) is only nonzero for very small $|\bm{k}_{2}-\bm{k}_{1}|$, this makes the integral difficult to perform numerically. In this section, we plan to develop the tools to deal with this issue. In Sec.~\ref{sec:accel}, we first formulate the expression of acceleration; then. in Sec.~\ref{sec:linear}, we simplify the expression to the one that can be directly used in numerical calculation.
%%%%%
%\subsection{Analytical expressions for the acceleration}
%\label{sec:accel}

From the scattering cross-section, we can express the pulsar kick acceleration originating from the recoil of the scattering as follows:
\begin{equation}
    \bm{a}(t)
    = -\frac{1}{M}\int (\bk_2-\bk_1) \frac{d\Omega_1}{4\pi}
    \frac{1}{\pi R^2}\frac{dL}{dE_1} dE_1 d\sigma
\end{equation}
Here, $M$ and $R$ represent the mass and the radius of the PNS and $d\Omega_1$ is the angular integration with respect to $\bk_1$.  We note that $L$ is the \textit{observed number luminosity} of neutrinos emitted from the PNS\@.
Thus, our estimate includes the contribution of the emitted neutrinos only quantified by $L$.  When neutrinos are rescattered and reabsorbed in the PNS, they have no net effect to the acceleration, but our formula involving $L$ has no such contribution by construction.
In the above formula, we can safely neglect the time dependence in $M$, while we treat $R(t)$ as a time-dependent quantity.
Plugging the explicit form of $d\sigma$ into this expression, we then reach
\begin{align}
    & \bm{a}(t) = -\frac{9G_F^2}{64\pi^4 MR(t)^2} \int d\Omega_1 d\Omega_2 dE E^2 \nl
    &\quad \times [1+\cos\theta_{1}\cos\theta_{2}-\sin\theta_{1}\sin\theta_{2}\cos(\varphi_{1}-\varphi_{2})]\nl
    &\quad \times (\bk_2-\bk_1)\, |\tilde{j}_{5,e}^{z}(\bm{k}_{2}-\bm{k}_{1},t)|^{2} \, \frac{dL(t)}{dE}\,,
    \label{eq:a_pk}
\end{align}
where we denoted the energy as $E=E_1=E_2$.  For the concrete estimate for the luminosity, for simplicity, we assume a simple relation between the number luminosity $L$ and the energy luminosity $L_E$ (the latter can be deduced from Ref.~\cite{Pons1999}) as
\begin{equation}
    \frac{dL(t)}{dE} = \frac{L_E(t)}{\langle E_\nu(t)\rangle}\, \delta(E-\langle E_\nu(t)\rangle)\,,
    \label{eq:L}
\end{equation}
where $\langle E_\nu(t)\rangle$ is the mean neutrino energy deduced from the temperature at the neutrino sphere.

%To appreciate this, recall that $L_E(t)$ and $\left<E_{\nu}(t)\right>$ are inputs from \cite{Pons1999}.  In that paper, the authors considered the neutrino re-absorption and re-scattering, but did not consider the scattering of neutrinos on the axial current. Hence, what we are doing here is equivalent to using their models as the 0th order approximation, and applying perturbation calculation to add the effect of the axial current on top of them. As long as the amount of neutrinos scattered by the axial current is considerably smaller than the total amount of emitted neutrinos (which is of course true since only 1\% of the total $\sim 10^{53} \; \mathrm{erg/s}$ neutrino luminosity is required to generate the observed kick velocity \cite{Maruyama2011,Nakazato2013}), this perturbation calculation would give a reasonably accurate result.

We substitute Eqs.~\eqref{eq:iso_and_aniso} and \eqref{eq:L} into Eq.~\eqref{eq:a_pk} and find that the acceleration can be separated into the 0-th order part and the 1-st order part with respect to $\alpha_1$ as $\bm{a}(t)=\bm{a}_{(0)}(t)+\bm{a}_{(1)}(t)$, where the 0-th order part is vanishing due to symmetry.
%\begin{align}
%    \bm{a}_{(0)}(t) &= -\frac{9G_{F}^{2}\langle E_{\nu}(t)\rangle L_{E}(t)}{64\pi^{4}MR(t)^2} \int d\Omega_1 d\Omega_2 \nl
%    & \times [1+\cos\theta_{1}\cos\theta_{2}-\sin\theta_{1}\sin\theta_{2}\cos(\varphi_{1}-\varphi_{2})] \nl
%    & \times (\bk_{2}-\bk_{1})\, |\tilde{j}_{5,e}^{z}(|\bk_{2}-\bk_{1}|,t)|^{2} \,,
%    \label{eq:a_pk_iso}
%\end{align}
%and the 1-st order part is
%\begin{align}
%    \bm{a}_{(1)}(t) &= -\alpha_1 \frac{9 G_{F}^{2}\langle E_{\nu}(t)\rangle L_{E}(t)}{32\pi^{4}MR(t)^2} \int d\Omega_1 d\Omega_2 \nl
%    & \times [1+\cos\theta_{1}\cos\theta_{2}-\sin\theta_{1}\sin\theta_{2}\cos(\varphi_{1}-\varphi_{2})] \nl
%    & \times (\bk_2-\bk_1)\, |\tilde{j}_{5,e}^{z}(|\bk_{2}-\bk_{1}|,t)|^{2} \frac{(\bk_2-\bk_1)_{z}}{|\bk_2-\bk_1|}\,.
%    \label{eq:a_pk_aniso}
%\end{align}
%Here, it should be noted that $|\bk_1|=|\bk_2|=k=\langle E_\nu(t)\rangle$.  Now it would be more convenient to replace $\bk_2$ by $\delta\bk=\bk_2-\bk_1$.  To this end, we can rewrite $d\Omega_2$ into the form of $d^3\bk_2$ which can be changed by $d^3\delta\bk$.  Then, we arrive at
%\begin{equation}
%    \int d\Omega_1 d\Omega_2 = \frac{2}{k}\int d\Omega_1 d^3\delta\bk\, \delta(\delta k^2 - 2\delta\bk\cdot\bk_1)\,.
%    \label{eq:vchange}
%\end{equation}

% approximation
To further simplify ${\bm a}_{(1)}(t)$, we introduce an approximation that works for $|\delta\bk|=|\bk_2-\bk_1|\sim 10^{-17}\,\text{MeV} \ll |\bk_1|=|\bk_2|=\langle E_\nu(t)\rangle$.
Then, we can drop higher order terms in $|\delta \bk|/|\bk_{1,2}|$.
%Then, we can approximate the integrand as
%\begin{equation}
%  \begin{split}
%    & 1+\cos\theta_1\cos\theta_2-\sin\theta_1\sin\theta_2\cos(\varphi_2-\varphi_1) \\
%    & \qquad\qquad\qquad = \frac{\delta k^2}{2k^2} + 2\cos\theta_1\cos\theta_2 \simeq 2\cos^2\theta_1\,.
%  \end{split}
%\end{equation}
%Also, in Eq.~\eqref{eq:vchange}, we can evaluate $\delta(\delta k^2-2\delta\bk\cdot\bk_1)\simeq \delta(2\delta\bk\cdot\bk_1)$ and so $\delta\bk\perp \bk_1$.
%With these preparations, we can express the acceleration as
%\begin{align}
%    a_{(0)}^z(t) &= -\frac{9G_{F}^{2}\langle E_{\nu}(t)\rangle L_{E}(t)}{64\pi^{4}MR(t)^2} \frac{1}{k^2}\int d\Omega_1\, d\delta k\, d\varphi_\delta\, \delta k \nl
%    & \quad \times 2\cos^2\theta_1 \delta k \cos\varphi_\delta \sin\theta_1 |\tilde{j}_{5,e}^{z}(\delta k,t)|^2\,,
%    \label{eq:a_pk_iso1}
%\end{align}
%which is zero after the integration with respect to $\varphi_\delta$.  We note that in our variable convention $\delta k^z=\delta k \cos\varphi_\delta \sin\theta_1$.
%In the same way, we can express the 1-st order acceleration as follows:
%\begin{align}
%    a_{(1)}^z(t) &= -\alpha_1\frac{9 G_{F}^{2}\langle E_{\nu}(t)\rangle L_{E}(t)}{32\pi^4 MR(t)^2} \frac{1}{k^2}\int d\Omega_1\, d\delta k\, d\varphi_\delta\, \delta k\nl
%    & \quad \times 2\cos^2\theta_1 \delta k \cos^2\varphi_\delta \sin^2\theta_1 |\tilde{j}_{5,e}^{z}(\delta k,t)|^2 \nl
%    & = -\alpha_1\frac{3G_{F}^{2}\langle E_{\nu}(t)\rangle L_{E}(t)}{10\pi^2 MR(t)^2} \frac{1}{k^2}\int d\delta k\, \delta k^2 |\tilde{j}_{5,e}^{z}(\delta k,t)|^2\,.
%    \label{eq:a_pk_aniso1}
%\end{align}
After some algebraic procedures, we find:
\begin{align}
    a_{(1)}^z(t) = -\alpha_1\frac{3G_{F}^{2}\langle E_{\nu}(t)\rangle L_{E}(t)}{10\pi^2 MR(t)^2} \frac{1}{k^2}\int d\delta k\, \delta k^2 |\tilde{j}_{5,e}^{z}(\delta k,t)|^2\,.
    \label{eq:a_pk_aniso1}
\end{align}
It is straightforward to confirm $({\bm a})^{x,y}(t)=0$, which is understood from axial symmetry around the $z$-axis.

In summary, introducing a dimensionless variable $\rho$ by $\delta k=\langle E_\nu(t)\rangle \rho$, we can write down the total acceleration as
\begin{align}
    & \bm{a}(t) = \hat{e}_z\, a_{(1)}^z(t) \nl
    &= -\hat{e}_z\, \alpha_1 \frac{3G_F^2 \left<E_{\nu}(t)\right>^2 L_E(t)}{10\pi^2 M R(t)^2} \int_0^\infty \!\! d\rho\, \rho^2 |\tilde{j}_{5,e}^z(\langle E_\nu(t)\rangle\rho,t)|^2\,.
    \label{eq:final_apk}
\end{align}
This expression for the kick acceleration is the central result in this work.
\vspace{1em}

%%%%%%%%%%
\paragraph*{Numerical results:}

Now, we should perform the $\rho$ integration in Eq.~\eqref{eq:final_apk} numerically and estimate the pulsar kick velocity quantitatively.

% magnetic field
We parametrized the magnetic field strength by $B_0$ and in this work we choose $B_0 = 10^{12} \unit{T}$.  Although this value is relatively high as compared to the standard NS strength, the magnetic field of the PNS is typically large and this value is rather a conservative choice in pulsar kick models driven by the magnetic field; see
Refs.~\cite{Kusenko1996,Lai1998,Maruyama2011,Adhya2017,Bhatt2017,Ayala2018,Yamamoto2023-1} for related works with comparable $B_0$.

\begin{figure}
    \includegraphics[width=0.95\columnwidth]{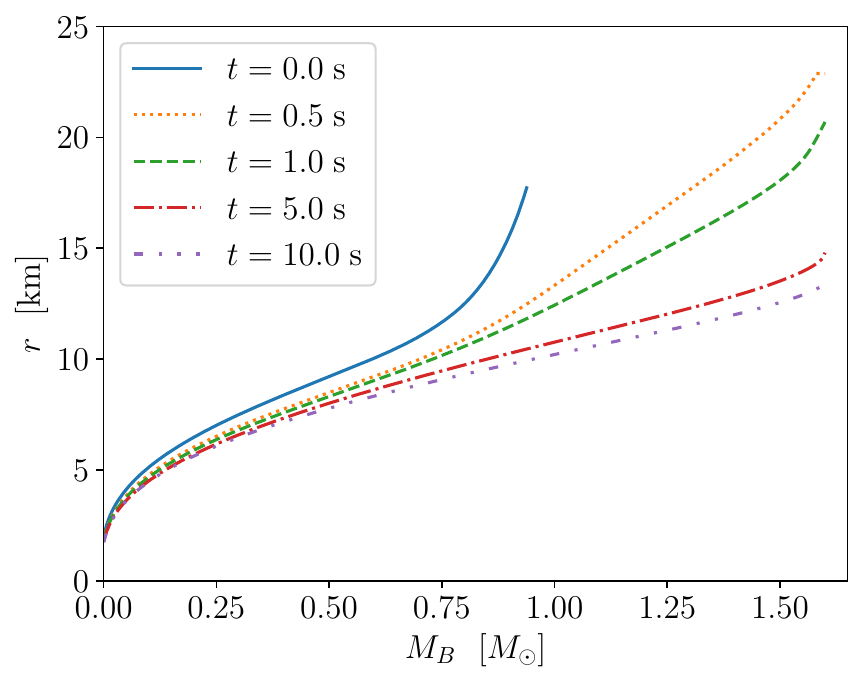}
    \caption{Relation between the enclosed baryonic mass, $M_B$, and the radius $r$ corresponding to the results in Ref.~\cite{Pons1999}.}
    \label{fig:Mr}
\end{figure}

To perform the numerical calculations, we need to specify the PNS model.  In this work we adopt results from Ref.~\cite{Pons1999} which presents the results in terms of not the radial distance $r$ but the enclosed baryonic mass defined by
$\frac{dM_B(r)}{dr} = 4\pi r^2 e^\lambda n_B(M_B) m_N$,
where $m_N$ is the nucleon mass and $e^{-\lambda}=\sqrt{1-2GM(r)/r}$ with the gravitational mass $M(r)$.  Since we are interested in the order estimate, we simplify the analysis by approximating $M(r)\simeq M_B(r)$.  Then, we can numerically solve $r(M_B)$ using given $n_B(M_B)$.  Figure~\ref{fig:Mr} shows $r(M_B)$ corresponding to the results in Ref.~\cite{Pons1999}.

%--- figure ---%
%--- figure ---%

\begin{figure}
    \includegraphics[width=0.95\columnwidth]{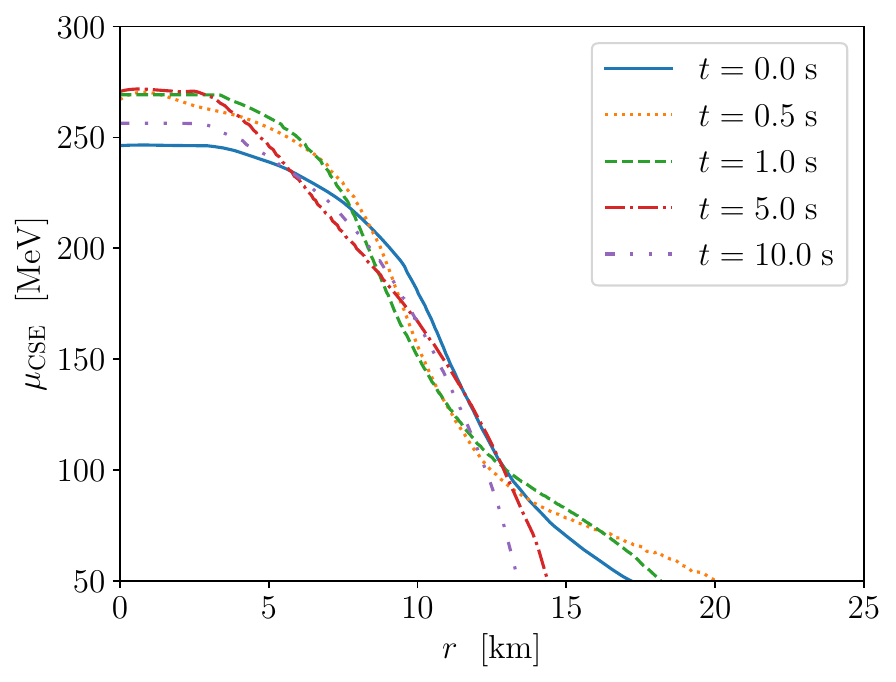}
    \caption{Coefficient in the chiral separation effect or the effective chemical potential as a function of the radial distance $r$ from the center of the star.}
    \label{fig:muCSE}
\end{figure}

In this way, we can translate $n_B(M_B)$ into $n_B(r)$.  The electron fraction, $Y_e$, can be deduced from Ref.~\cite{Pons1999} as well, and the spatial profile of the corresponding chemical potential, $\mue(r)$, is obtained.  Since the electron mass is tiny as compared to the typical scale of $\mue(r)$, the coefficient in the chiral separation effect, $\muCSE$ in Eq.~\eqref{eq:CSE} is almost identical with $\mue$ at $T=0$, and the spatial profile including the finite temperature effects can be estimated from Eq.~\eqref{eq:CSE} as presented in Fig.~\ref{fig:muCSE}.

%--- figure ---%
%--- figure ---%

We now perform the Fourier transformation of $\bj_{5,e}=(e/2\pi^2)\muCSE \bB$ and numerically estimate $|\tilde{\bj}_{5,e}(k)|$ which damps quickly at $k\sim 1/(10\text{km})\sim 10^{-17}\unit{MeV}$ as mentioned before.  The Fourier transformed current is dimensionless, and for such a spatial profile with a typical extension of $\sim 10\unit{km}$, it turns out that $|\tilde{\bj}_{5,e}(0)|\sim 10^{54}$.  This gigantic number arises from the macroscopic scale of the star radius.  It would be an instructive check to make the order estimate at this point.  From Ref.~\cite{Pons1999} we see that $\langle E_\nu(t)\rangle \sim 10\unit{MeV}$ and $L_E(t)\sim 50\times 10^{51}\unit{erg/s}$ in small $t$ regions.  Then, let us approximate that $|\tilde{j}_{5,e}^z(\langle E_\nu(t)\rangle\rho)|^2$ has a support for $0\le \rho\le 10^{-18}$ with a constant amplitude $\sim 10^{54}$.  In the denominator $M=1.6M_\odot$ and $R\sim 10\unit{km}$ would be a reasonable choice.  Under these assumptions Eq.~\eqref{eq:final_apk} leads to $a\sim \alpha_1\times 5000\unit{km/s^2}$.

We still need to fix the value or at least the order of $\alpha_1$.  To the best of our knowledge, there is no consensus about $\alpha_1$ in the PNS simulations.  This is partially because the anisotropy predicted by the supernova explosion simulation has large ambiguity (see Refs.~\cite{Nakamura2019, Burrow2020}), and also because the evolution of the anisotropy within the short time scale less than one minute after the explosion has not been thoroughly studied.  In some supernova explosion simulations~\cite{Scheck2006,Nordhaus2010}, anisotropy of the ejecta at $10\%$ level is reported, and if it scales with the anisotropy of the core, it would be sensible to constrain $|\alpha_1| > 0.1$.  We note that other studies~\cite{Lai1998, Yamamoto2020} show highly anisotropic $\mue(\br,t)$, $\bB(\br,t)$, and $T(\br,t)$, which could persist during the first seconds of the newly born NSs.  The PNSs are held by the supernova remnant at the first seconds of their life (see a review~\cite{Enoto2019} for more details), namely, a cluster of hot and highly inhomogeneous plasma, which could distort the magnetic field distribution.  Furthermore, in NS cores, the
instability mechanisms amplifying the local magnetic field strength could also favor the initial anisotropy.  All these arguments suggest an even larger value of $|\alpha_1|$.

\begin{figure}
    \includegraphics[width=0.95\columnwidth]{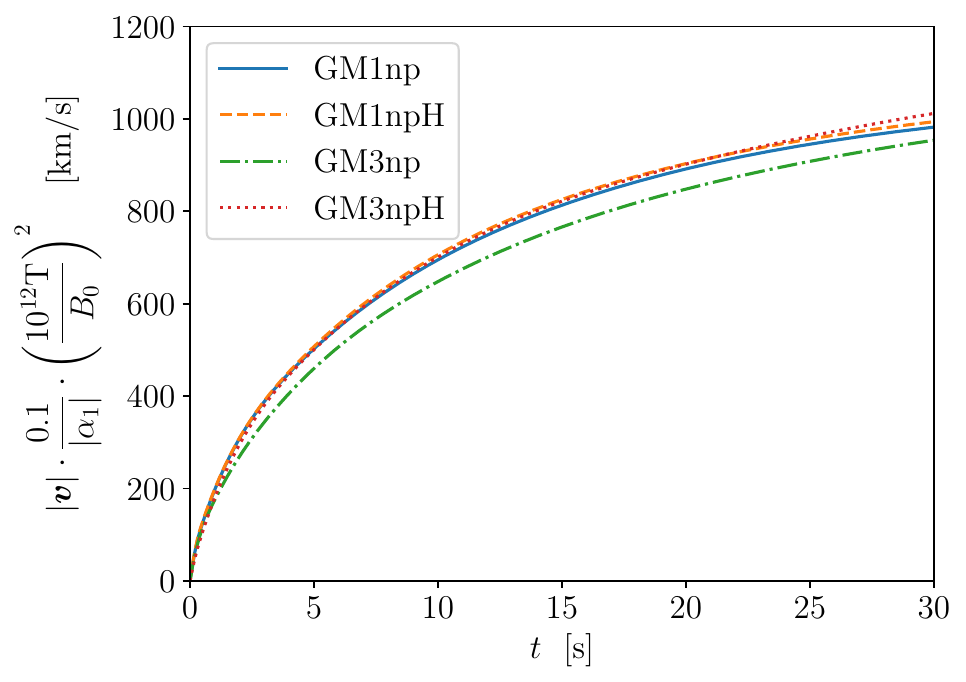}
    \caption{Velocity as a result of integrated acceleration by the chiral pulsar kick as a function of time.}
    \label{fig:v}
\end{figure}

% kick velocity
We are ready to proceed to the numerical calculations using the date from Ref.~\cite{Pons1999}.  Because we assume constant $B_0$ and $\alpha_1$, the final results are simply proportional to $\alpha_1 B_0^2$, which is obvious from Eqs.~\eqref{eq:j5_e0} and \eqref{eq:final_apk}.  We finally get the velocity as shown in Fig.~\ref{fig:v}.  Here, GM1 and GM3 refer to the stiff and the soft equation-of-state models in Ref.~\cite{1991PhRvL..67.2414G}, respectively, and np represents the model with nucleonic degrees of freedom only and npH represents the model that allows for the hyperon degrees of freedom.

%--- figure ---%
%\begin{figure}
%    \includegraphics[width=0.95\columnwidth]{fig_a.pdf}
%    \caption{Acceleration by the chiral pulsar kick as a function of time.}
%    \label{fig:a}
%\end{figure}
%--- figure ---%

%--- figure ---%

From the figure, we see that the pulsar kick behavior of different models stays close to each other: the velocity first rises rapidly and at around $t \gtrsim 30 \unit{s}$, the growing curve is saturated around
\begin{equation}
    |\bm{v}_{\text{kick}}| \approx \frac{|\alpha_1|}{0.1}\biggl(\frac{B_0}{10^2\unit{T}}\biggr)^2 \times 1000\unit{km/s}\,.
\end{equation}
As for the direction of the kick velocity, it depends on both the magnetic field and the anisotropic profile in momentum space, and in the present simple setup, the velocity is positive in the direction anti-parallel to the magnetic field for positive $\alpha_1$.
\vspace{1em}

%%%%%%%%%%
\paragraph*{Summary:}

We proposed a mechanism which effectively converts anisotropy to propulsion through scattering between neutrinos and the background axial current $\bj_{5,e}$ of electrons.  The presence of $\bj_{5,e}$ is a robust consequence from the chiral separation effect, but constant $\bj_{5,e}$ does not discriminate parallel/anti-parallel scatterings along the magnetic axis.  We find, however, that, once momentum space anisotropy in either density or magnetic profile is coupled, scattering involving $\bj_{5,e}$ plays a significant role for the chiral anisotropy conversion that transforms anisotropy to acceleration. 

In the past, the anisotropy effect and the magnetic effect were separately considered, and our mechanism is a hybrid one that has been overlooked.  Interestingly, our estimate concluded that the resulting pulsar kick can be strong enough to explain the velocity of order of $\sim 1000\unit{km/s}$ if the PNSs have $B_0\sim 10^{12}\unit{T}$ and $\sim 10\%$ momentum anisotropy.  Therefore, our mechanism should be one of major components and it should be taken into account for quantitative analyses together with others.  We shall make an important remark that the acceleration direction in our proposed scenario is determined by not only $\bB$ but also the shape of $\alpha_1$ once nonuniformity of $\alpha_1$ is considered.  This could also explain the observation that the velocity is not perfectly aligned to the magnetic direction.

There are several future extensions.  In this work we neglected the nucleon contributions because the CSE for nucleons should be suppressed by mass.  Nevertheless, it is technically easy to include the nucleon contributions and we already found that the correction is up to a few $\%$.  We will report this in details elsewhere.  Another more ambitious direction is to develop the PNS simulation fully taking account of the background $\bj_{5,e}$ effect.  For the order estimate as demonstrated here, outputs from Ref.~\cite{Pons1999} suffice for the purpose.  However, the realistic equation-of-state model is better constrained nowadays.  Besides, we simply assumed the strength of anisotropy, but the CSE would work in favor of forming anisotropic distributions.  To implement such backreactions, we should go beyond the current scope, i.e., the chiral radiation hydrodynamics,  which would deserve future investigations.

%%%%%%%%%%
\begin{acknowledgments}
  The authors would like to thank Alejandro~Ayala, David~Blaschke and Matthias~Kaminski for discussions at the ECT* workshop, ``Stronly Interacting Matter in Extreme Magnetic Fields.''
  The authors are also grateful to Naoki~Yamamoto and Di-Lun~Yang for useful comments.
  This work was partially supported by JSPS KAKENHI Grant Nos.\ 22H01216 (K.F.) and 22H05118 (K.F.).
\end{acknowledgments}

\bibliographystyle{apsrev4-1}
\bibliography{pkick.bib}

\end{document}